\documentclass[twocolumn,showpacs]{revtex4}
\usepackage[xdvi]{graphicx}%
\usepackage{dcolumn}
\usepackage{amsmath,amssymb}

\begin{document}

\title{Zon-Cohen singularity and negative inverse temperature in a trapped particle limit}

\author{Takahiro Nemoto}
\affiliation{
Department of Basic Science,
The University of Tokyo, Tokyo, 153-8902, Japan and\\
Department of Physics, Kyoto University, Kyoto 606-8502, Japan}
\date{\today}

\begin{abstract}
We study a Brownian particle on a moving periodic potential.
We focus on the statistical properties of the work done by the potential
and the heat dissipated by the particle.
When the period and the depth of the potential are both large, by using a boundary layer analysis, we
calculate a cumulant generating function
and
a biased distribution function.
The result allows us to understand
a Zon-Cohen singularity
for an extended fluctuation theorem
from a view point of rare trajectories characterized by a negative inverse temperature
of the biased distribution function.

\end{abstract}


\maketitle



\section{Introduction}



In 1993, the fluctuation theorem was discovered \cite{Evans1993}.
The theorem claims a symmetry property of the fluctuation of entropy production
and provides us with a deep understanding
of nonequilibrium physics \cite{Gallavotti_cohen, Kurchan_fluctuation, Maes_fluctuation, Crooks_fluctuation, Lebowitz_Spohn}.
The first verification of the theorem in real experiments was done by
Wang {\it et al} in 2001 \cite{Wang}. 
They used a Brownian particle dragged by an optical tweezer,
and checked the fluctuation theorem for the work done by the tweezer.
See also Ref. \cite{Zon-Cohen1}
for the detailed analysis of the system.
In the stationary state,
the expectation values of 
the work done by the tweezer and the heat dissipated by the
particle are equal to each other.
However, Zon and Cohen \cite{Zon-Cohen2, Zon-Cohen3} predicted that the fluctuations
of the work and the heat
were different.
They pointed out that
the fluctuation theorem for the heat was modified,
while the fluctuation theorem for the work was valid.
They called the modified relation an extended
fluctuation theorem.
See Refs. \cite{Garnier, Bonetto, Baiesi1, Visco, Harris, Rakos, Puglis, Noh-Park}
for recent studies of the extended fluctuation theorem.

In order to examine the extended fluctuation theorem,
let us consider a cumulant generating function,
\begin{equation}
G(h)=\frac{1}{t} \log \left \langle e^{hQ(t)} \right \rangle,
\label{Introduction1}
\end{equation}
and a biased distribution function, 
\begin{equation}
P_h(x,t)=e^{-t G(h)}\left \langle \delta (x(t)-x)e^{hQ(t)}\right \rangle,
\label{introduction2}
\end{equation}
where
$Q(t)$ is the accumulated heat from $t=0$ to $t=t$
and $x(t)$ is the position of the particle at time $t$.
The parameter $h$ is called a biasing field,
because one may understand that
 the right-hand side
of (\ref{introduction2})
is an expectation value of $\delta(x(t)-x)$
with respect to
a path probability
given by multiplying
the original path probability by a biasing
factor $e^{hQ(t)-tG(h)}$.
A hardly measurable trajectory, from which we evaluate a large value
of $Q(t)$, has
a larger weight in (\ref{introduction2}) with $h>0$ than
in the original distribution function 
[(\ref{introduction2}) with $h=0$]. This indicates that
the biasing field is related to rare trajectories.
Indeed, the large deviation theory 
connects rare trajectories with biasing fields more directly \cite{Dembo, Touchette}.

The extended fluctuation theorem was equivalent to
a singularity of the cumulant generating function (\ref{Introduction1})
\cite{Zon-Cohen2, Zon-Cohen3}.
When $|h|$ is larger than a special value $h_c$,
$G(h)$ becomes singular.
In this paper, we call the singularity a Zon-Cohen singularity.
The fact that the Zon-Cohen singularity emerges
when the biasing field is larger than $h_c$ indicates
that the singularity is related to rare trajectories.
However, the relationship between the singularity
and the behavior of the particle
in hardly measurable trajectories
is still unclear.
Since the significance of fluctuation in nonequilibrium physics 
has been recognized recently, it is important to study
the relationship using a systematic method
and investigate whether
the same kind of singularity occurs
not only for the heat but also for the other quantities.

In this paper, we consider a Brownian particle on a moving periodic potential.
The model is the overdamped case of the model studied by Lebowitz and Spohn
in Ref. \cite{Lebowitz_Spohn}.
By using a boundary layer analysis,
we calculate a cumulant generating function
and a biased distribution function when
the period and the depth of the potential are both large.
As the result, we find that the biased distribution function
becomes a canonical distribution function, where the inverse temperature
is modified by $h$.
When $|h|>h_c$, the inverse temperature becomes negative and
the two limiting operations, which are
the trapped particle limit and a limit of large observation time,
become non-interchangeable.
This non-interchangeability corresponds to
the Zon-Cohen singularity.
We also check a conditional distribution function
given $Q(t)/t=q$.
It allows us to understand
how hardly measurable
trajectories cause the singularity.
The discussion might indicate that the same kind of singularity exists
in the other quantities.

The organization of the paper is the following.
In Section \ref{Model}, we define a model and
introduce a biased process.
In Section \ref{Sectionresult} and \ref{Sectionderivation},
we state main results
and derive them.
Finally, Section \ref{concluding_remarks} is devoted to concluding remarks.
The Boltzmann constant $k_{\rm B}$ is set to unity throughout the paper.


\section{Set up}
\label{Model}

\subsection{Model}
We consider a one-dimensional Brownian particle. 
The temperature of the solvent is denoted by $T$.
The position of the particle is denoted by 
$x(t)\in \bf \mathbb R$.
A force $-\partial U(x)/\partial x$ is exerted on the particle,
where $U(x)$ is a periodic potential.
The period of the potential is $2L$. That is, $U(x)$
satisfies
\begin{equation}
U(x)=U(x+2nL)
\end{equation}
for $n=\pm 1, \pm 2, \cdots$.
We move the potential with a constant velocity $v$ toward the negative direction of $x$. 
The motion of the particle is described by the Langevin equation
\begin{equation}
\dot x(t)=-\frac{1}{\gamma}\frac{\partial }{\partial y}U(y)\bigg{|}_{y=x(t) + v t} + 
\sqrt{\frac{2T}{\gamma}}\xi(t),
\label{x}
\end{equation}
where $\xi(t)$ is the Gaussian white noise that satisfies
$\left \langle \xi(t) \right \rangle=0$ and $\left \langle \xi(t) \xi(s) \right \rangle=
\delta (t-s)$, and $\gamma$ is a friction constant.
In order to make the analysis easy, we introduce a new variable $y(t)$
as the position of the particle  measured within a reference frame that moves with the
periodic potential.
Concretely, $y(t)$ is defined as
\begin{equation}
y(t)\equiv x(t)+vt - 2nL,
\label{y}
\end{equation}
where  $n$ is an integer determined by
$-L\leq x(t)+vt-2nL<L$.
Note that $y(t)$ is confined to $[-L,L)$. 
From (\ref{x}), we obtain the Langevin equation for $y(t)$ as 
\begin{equation}
\dot y(t)=-\frac{1}{\gamma} \frac{\partial}{\partial y}U\left (y(t) \right )+v  +\sqrt{\frac{2T}{\gamma}} \xi(t).
\label{y}
\end{equation}
This system can be realized in real experiments.
See, for example, Ref. \cite{Brownian_optical_tweezer}.
Recently, the system has been used for experimental tests
of some nonequilibrium relations \cite{Bechinger, Ciliberto,Toyabe}.

\begin{figure}[tbh]
\includegraphics[width=8.0cm]{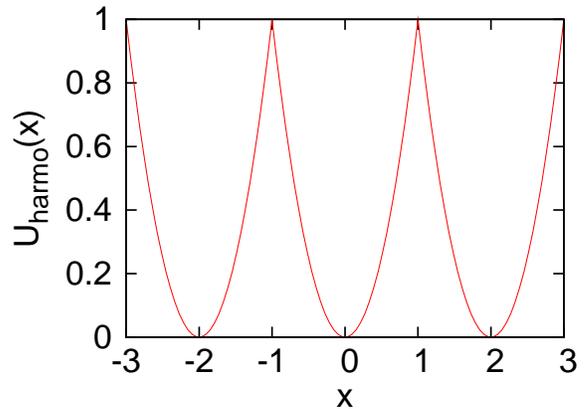}
\caption{(Color online)
A periodic potential that satisfies (\ref{potential_condition}).
We drew (\ref{harmonic}) by setting $k=2$ and $L=1$.
}
\label{fig4}
\end{figure}

We consider periodic potentials $U(x)$ that satisfy
the following condition
\begin{equation}
\lim_{L\rightarrow \infty}
\frac{\partial U(y)}{\partial y}\bigg |_{y=YL} = \infty
\label{potential_condition}
\end{equation}
for $0<|Y|\leq1$.
A harmonic potential 
\begin{equation}
U_{\rm harmo}(x)=\frac{1}{2}k(x-2nL)^2,
\label{harmonic}
\end{equation}
and a quartic potential
\begin{equation}
U_{\rm quart}(x)=\frac{1}{4}k_4(x-2nL)^4 ,
\end{equation}
where $n$ is an integer determined by an inequality $-L \leq x-2nL< L$,
are the examples that satisfy (\ref{potential_condition}).
We mention that a linear potential 
\begin{equation}
U_{\rm linear}(x)=k_{1}|x-2nL|
\end{equation}
does not satisfy (\ref{potential_condition}).
We display (\ref{harmonic}) in Fig. \ref{fig4}.


We consider the work $W(t)$ done by the periodic potential.
Since the periodic potential exerts the force $-\partial U(y)/\partial y | _{y=x(t)+vt}$ 
on the particle and
moves with the constant velocity $-v$,
we find that $\dot W(t)$ is calculated as
\begin{equation}
\dot W(t)= (- v)\left [- \frac{\partial }{\partial y}U(y)\bigg{|}_{y=x(t)+vt} \right ].
\label{W} 
\end{equation}
Next, we consider the heat $Q(t)$ dissipated by the particle. According to Sekimoto's argument \cite{Sekimoto}, the rate of the heat dissipation
is evaluated as
\begin{equation}
\dot Q(t)= \dot x \circ  \left [\gamma \dot x - \sqrt{2\gamma T}\xi(t) \right ],
\label{Q}
\end{equation}
where the multiplication $\circ$ represents the Stratonovich
interpretation \cite{Gardiner}.
We can immediately check that the first law of thermodynamics is satisfied.
That is,
\begin{equation}
\int _{t_1}^{t_2}dt \left ( \dot W(t)- \dot Q(t) \right )= U(x(t_2)+vt_2) -U(x(t_1)+vt_1).
\label{first_law_of_thermodynamics}
\end{equation}
We also express (\ref{W}) and (\ref{Q}) by using $y(t)$.
The result is
\begin{equation}
\dot W(t) = v\frac{\partial}{\partial y}U\left (y(t) \right ),
\label{W_y}
\end{equation}
\begin{equation}
\dot Q(t) = \frac{1}{\gamma} \left (\frac{\partial U(y)}{\partial y} \right )^2
-\sqrt{\frac{2T}{\gamma}}\frac{\partial U(y)}{\partial y} \circ \xi(t) .
\label{Q_y}
\end{equation}

We denote 
by $\left \langle \, \right \rangle_{p}$ the expectation value
over the noise $(\xi(s))_{s=0}^{\infty}$ with an 
initial distribution function $p(y_0)=\left \langle \delta(y(0)-y_0 )\right \rangle$. By using the notation, we define a
joint distribution function by
\begin{equation}
P(y_0,y,t|p)=\left \langle\delta(y(0)-y_0 ) \delta (y(t) - y)\right \rangle_{p}.
\end{equation}

\subsection{Biased process and cumulant generating functions}

We introduce a biased process.
We consider a function $f(t)$, which
depends on the trajectory of the particle $(y(s))_{s=0}^{t}$.
We define the expectation value of $f(t)$ in the biased process 
by
\begin{equation}
\begin{split}
&\left \langle f (t)\right \rangle^{h_w,h_q}_{p} \equiv e^{-tG(h_w,h_q,t|p)}
\left \langle f(t)e^{W(t)h_w+Q(t)h_q}\right \rangle_{p},
\label{biasedprocess}
\end{split}
\end{equation}
where $G(h_w,h_q,t|p)$ is
a cumulant generating function defined by
\begin{equation}
G(h_w,h_q, t|p)= \frac{1}{t}
\log 
\left \langle e^{W(t)h_w+Q(t)h_q}
\right \rangle _{p}.
\label{Definition_of_G}
\end{equation}
We note that the cumulant generating function corresponds to
a thermodynamic free energy
according to the thermodynamic formalism \cite{Ruelle}.
The two parameters $h_w$ and $h_q$ are
called biasing fields.
When we set $h_w=h_q=0$, 
the biased expectation value (\ref{biasedprocess})
returns to the original expectation value
and the cumulant generating function (\ref{Definition_of_G})
becomes $0$.
By using (\ref{biasedprocess}), we define a biased joint distribution function
as
\begin{equation}
\begin{split}
&P_{h_w,h_q}(y_0,y,t|p) =\left \langle \delta (y(t)-y)
\delta (y(0)-y_0)  \right \rangle^{h_w,h_q}_p.
\label{biased_distribution}
\end{split}
\end{equation}

In order to analyze the large $t$ behavior of the cumulant generating
function,
we define two types of functions.
The first one is 
\begin{equation}
G_{\rm scaled}(h_w,h_q)= \lim_{t \rightarrow \infty}G(h_w,h_q,t|p).
\label{usualdefinition_ofcumulant}
\end{equation}
The second one is
\begin{equation}
 H_{\rm ex}(h_w,h_q|p)  =\lim_{t\rightarrow \infty}
t \left [ G(h_w,h_q,t |p) - G_{\rm scaled}(h_w,h_q) \right ]  .
\label{Definition_of_excess_cumulant}
\end{equation}
(\ref{usualdefinition_ofcumulant}) is called a
scaled cumulant generating function \cite{Touchette}.
(\ref{Definition_of_excess_cumulant})
is an excess quantity of the cumulant generating function,
which was used for the calculation
of an excess heat in Ref. \cite{Sagawa-Hayakawa}.
We call it an excess cumulant generating function.
Since the excess cumulant generating function depends on the initial distribution function $p$, 
we explicitly indicated it in (\ref{Definition_of_excess_cumulant}).
By using (\ref{usualdefinition_ofcumulant}) and (\ref{Definition_of_excess_cumulant}), we may express 
$G(h_w,h_q,t|p)$ as 
\begin{equation}
G(h_w,h_q,t|p) \simeq G_{\rm scaled}(h_w,h_q) + \frac{1}{t} H_{\rm ex}(h_w,h_q|p).
\label{asymptotic_expression_of_cumulant}
\end{equation}
We note that the difference of the left-hand side and the right-hand side
of (\ref{asymptotic_expression_of_cumulant})
is $O(e^{-at})$,
where $a$ is a positive constant.
See (\ref{qasymptotics}).

\subsection{Biased distribution function and conditional distribution function}
Here, we show a useful relation between the
biased distribution function and a conditional distribution function.
Let us consider a joint distribution function of $y(0)$, $y(t)$,
and $Q(t)/t$, which is defined by
\begin{equation}
\begin{split}
&P(y_0,y,q,t|p) \\
&\equiv \left \langle \delta(y(0)-y_0)\delta(y(t)-y) \delta (Q(t)/t-q)
\right \rangle_{p}.
\end{split}
\end{equation}
By using this definition in the right-hand side of (\ref{biased_distribution}), we
obtain a relation
\begin{equation}
\begin{split}
&\log P_{0,h_q}(y_0,y,t|p) 
+t G(0,h_q,t|p)\\
&= \log  \int dq P(y_0,y,q,t|p)e^{t q h_q}.
\label{relationjointandbiase}
\end{split}
\end{equation}
Then, we define a function $I(y_0,y,q,t|p)$ by
\begin{equation}
I(y_0,y,q,t|p)\equiv -\frac{1}{t}\log P(y_0,y,q,t|p).
\label{definitionI}
\end{equation}
For $I(y_0,y,q,t|p)$, we assume the following asymptotic form
\begin{equation}
\begin{split}
&I(y_0,y,q,t|p)= I_0(q) +\frac{1}{t}I_1(y_0,y,q|p)+o(1/t)
\label{asymptotic_form}
\end{split}
\end{equation}
when $t$ is large.
(\ref{asymptotic_form}) corresponds to a large deviation property,
and $I_0(q)$ is a large deviation function for
$Q(t)/t$.
By using the asymptotic form and a saddle point method,
we calculate the right-hand side of (\ref{relationjointandbiase}).
The result is
\begin{equation}
\begin{split}
&t \max_{q}\left [h_q q -I_0(q)\right ] -I_1(y_0,y,q^*|p)+\frac{1}{1/t}o(1/t),
\label{asymptoticlarge}
\end{split}
\end{equation}
where $q^{*}$ is defined as
\begin{equation}
q^*\equiv \underset{q}{\rm argmax}\left [h_q q -I_0(q)\right ].
\label{q*condition}
\end{equation}
From (\ref{asymptotic_expression_of_cumulant}),
(\ref{relationjointandbiase}), and (\ref{asymptoticlarge}),
we thus obtain
\begin{equation}
G_{\rm scaled}(0,h_q)= \max_{q}\left [h_q q -I_0(q)\right ],
\label{scaledlegendre}
\end{equation}
and
\begin{equation}
\begin{split}
&\log P_{0,h_q}(y_0,y,t|p) + H_{\rm ex}(0,h_q|p)\\
&=-I_1(y_0,y,q^*|p)+\frac{1}{1/t}o(1/t).
\label{secondlegendre}
\end{split}
\end{equation}
Here, (\ref{scaledlegendre}) is a well-known relation between
a large deviation function and a scaled cumulant generating function \cite{Dembo,Touchette}.
By combining (\ref{secondlegendre}) with (\ref{asymptotic_form})
and noticing the normalization condition for $P_{0,h_q}(y_0,y,t|p)$,
we arrive at
\begin{equation}
P_{0,h_q}(y_0,y,t|p) =\frac{P(y_0,y,q^*,t|p)}{P(q^*,t|p)}
+\frac{1}{1/t}o(1/t),
\label{equivalence}
\end{equation}
where $P(q^*,t|p)$ is a normalization constant defined by
\begin{equation}
P(q^*,t|p)\equiv \int dy_0 dy P(y_0,y,q^*,t|p) .
\end{equation}
Since $P(q,t|p)$ is the distribution function of $Q(t)/t$, 
we find that the biased distribution function
is nothing but the conditional distribution function
of $y(0)$ and $y(t)$ given
$Q(t)/t=q^*$.


\section{Results}
\label{Sectionresult}
We denote the stationary distribution function
of $y$ by $p_{\rm st}^{U,v,\beta}(y)$,
where the superscripts $U$, $v$, and $\beta$ indicate the periodic potential,
the moving velocity of the potential, and the inverse temperature of the solvent,
respectively. 
Then, we define a canonical distribution function $p_{\rm can}^{U,v,\beta}(y)$
by
\begin{equation}
p_{\rm can}^{U,v,\beta}(y) \equiv \frac{1}{Z(v,\beta)}e^{-\beta U(y)+\gamma v \beta y},
\label{canonicaldistribution}
\end{equation}
where $Z(v,\beta)$ is a normalization constant determined by 
\begin{equation}
Z(v,\beta) = \int _{-L}^{L}dye^{-\beta U(y)+\gamma v \beta y}.
\end{equation}
The first result is that $p_{\rm st}^{U,v,\beta}(y)$ approaches $p_{\rm can}^{U,v,\beta}(y)$
when $L$ is large:
\begin{equation}
p_{\rm st}^{U,v,\beta}(y)\sim p_{\rm can}^{U,v,\beta}(y)
\label{result1}
\end{equation}
for $y=YL$ ($-1<Y<1$). The definition of the symbol $\sim$
is the following.
For functions $A(Y)$ and $B(Y)$,
which depend on $L$,
we define $A(Y)\sim B(Y)$ for $-1<Y<1$ as $\lim _{L \rightarrow \infty}
(1/L)\log [A(Y)/ B(Y)] =0$ for each fixed $Y$. We use the symbol $\sim$ throughout the paper.

The second result is about the scaled cumulant generating function $G_{\rm scaled}(h_w,h_q)$.
The function always becomes a quadratic function 
in the limit $L\rightarrow \infty$. That is, 
\begin{equation}
\lim_{L \rightarrow \infty}G_{\rm scaled}(h_w,h_q)=\gamma v^2
(h_w+h_q)+T\gamma v^2(h_w+h_q)^2.
\label{scaledlimit}
\end{equation}
It should be stressed that the result is always valid whenever
the periodic potential satisfies (\ref{potential_condition}).
From (\ref{asymptotic_expression_of_cumulant}) and (\ref{scaledlimit}), we notice that 
\begin{equation}
\begin{split}
&\lim_{L \rightarrow \infty} \lim_{t \rightarrow \infty}
G(h_w,h_q,t|p)\\
&=\gamma v^2
(h_w+h_q)+T\gamma v^2(h_w+h_q)^2
\label{taufirst}
\end{split}
\end{equation}
for any $h_w$ and $h_q$.
The system under consideration was analyzed in Ref \cite{Lebowitz_Spohn} 
by Lebowitz and Spohn. They proved the fluctuation theorem
in this system.
The theorem is written as
\begin{equation}
G_{\rm scaled}(h_w,h_q)=G_{\rm scaled}(-h_w,-\beta-h_q)|_{v\rightarrow -v},
\label{fluctuationtheorem}
\end{equation}
where $|_{v\rightarrow -v}$ means that the sign of $v$ is reversed.
We will re-derive (\ref{fluctuationtheorem}) in Section \ref{Sectionderivation} for
the sake of completeness.
From (\ref{fluctuationtheorem}), we obtain  
\begin{equation}
\begin{split}
&\lim_{L\rightarrow \infty} \lim_{t \rightarrow \infty}G(h_w,h_q,t|p)
\\
&=\lim_{L\rightarrow \infty} \lim_{t \rightarrow \infty}G(-h_w,-\beta - h_q,t|p)|_{v\rightarrow -v},
\label{fluctuationtheorem1}
\end{split}
\end{equation}
which can also be verified by using (\ref{taufirst}) directly.

Hereafter, we focus on the case in which
the initial distribution function $p(y)$ is equal to a stationary
distribution function $p_{\rm st}^{U,v^{\prime},\beta^{\prime}}(y)$,
where $\beta^{\prime}(>0)$ and $v^{\prime}$
represent an inverse temperature and a velocity in another system.
The third result is about the behavior of the biased joint distribution function 
$P_{h_w,h_q}(y_0,y,t|p_{\rm st}^{U,v^{\prime},\beta^{\prime}})$ when 
$t$ and $L$ are both large. That is,
\begin{equation}
\begin{split}
&P_{h_w,h_q}(y_0,y,t|p_{\rm st}^{U,v^{\prime},\beta^{\prime}})\\
&\sim p_{\rm can}^{U,v_{\rm i},\beta_{\rm i}}(y_0)p_{\rm can}^{U,v_{\rm f},\beta_{\rm f}}(y) 
+O(e^{-at})
\label{biased_result_limit0}
\end{split}
\end{equation}
for $y_0=Y_0L$, $y=YL$ $(-1<Y_0,Y<1)$ with
\begin{equation}
\beta_{\rm i}= \beta^{\prime} -h_q,
\label{modified_temp_ini}
\end{equation}
\begin{equation}
v_{\rm i}= \frac{v^{\prime}\beta^{\prime}+v(h_q+h_w)}{\beta^{\prime}-h_q},
\label{modified_velo_ini}
\end{equation}
\begin{equation}
\beta_{\rm f}= \beta  + h_q,
\label{modified_temp_fin}
\end{equation}
\begin{equation}
v_{\rm f}=\frac{v(\beta+h_q+h_w)}{\beta +h_q}.
\label{modified_velo_fin}
\end{equation}


From (\ref{modified_temp_ini}) and (\ref{modified_temp_fin}),
we notice that the inverse temperatures
of the canonical distribution functions
in (\ref{biased_result_limit0})
can become
negative values.
It turns out that the excess cumulant generating function
has different asymptotic behaviors according to whether the inverse temperature
is negative or not.
This is the fourth result. Concretely, when we set $L$ to be large,
the excess cumulant generating function
$H_{\rm ex}(h_w,h_q|p_{\rm st}^{U,v^{\prime},\beta^{\prime}})$
satisfies
\begin{equation}
H_{\rm ex}(h_w,h_q|p_{\rm st}^{U,v^{\prime},\beta^{\prime}}) 
=O(U(L))
\label{asymptoticH1}
\end{equation}
for $\beta_i<0$ or $\beta_f<0$, and
\begin{equation}
H_{\rm ex}(h_w,h_q|p_{\rm st}^{U,v^{\prime},\beta^{\prime}}) 
=O(1)
\label{asymptoticH2}
\end{equation}
for $\beta_i>0$ and $\beta_f>0$.
From (\ref{asymptotic_expression_of_cumulant}), (\ref{asymptoticH1}), and (\ref{asymptoticH2}), we find
\begin{equation}
\begin{split}
&\lim_{t \rightarrow \infty} \lim_{L\rightarrow \infty} G(h_w,h_q,t|p)
=\infty
\label{Lfirst}
\end{split}
\end{equation}
for $\beta_i<0$ or $\beta_f<0$, and
\begin{equation}
\begin{split}
&\lim_{t \rightarrow \infty} \lim_{L\rightarrow \infty} G(h_w,h_q,t|p)
=\lim_{L\rightarrow \infty} \lim_{t \rightarrow \infty}  G(h_w,h_q,t|p)
\label{taufirst2}
\end{split}
\end{equation}
for $\beta_i>0$ and $\beta_f>0$.
(\ref{taufirst2}) shows that
the two limiting operations, which are $L\rightarrow \infty$
and $t\rightarrow \infty$, are interchangeable
and the symmetry property of the fluctuation theorem is satisfied
when $\beta_{\rm i}$ and $\beta_{\rm f}$ are both positive.
However, when $\beta_{\rm i}$ or $\beta_{\rm f}$ is negative,
the two limiting operations
become non-interchangeable.
If we take $t \rightarrow \infty$ first, the
cumulant generating function satisfies
the fluctuation theorem (\ref{fluctuationtheorem1}).
On the other hand,
if we take $L\rightarrow \infty$ first,
the cumulant generating function diverges as shown in (\ref{Lfirst}).
This divergence
corresponds to the Zon-Cohen singularity \cite{Zon-Cohen2, Zon-Cohen3}.

\subsection{The negative inverse temperature
and the Zon-Cohen singularity}
By substituting the explicit expression of
the scaled cumulant generating function (\ref{fluctuationtheorem})
into (\ref{q*condition}) and (\ref{scaledlegendre}),
we obtain a relation between $q^*$ and $h_q$ as
\begin{equation}
h_q=\frac{q^*-\gamma v^2}{2T\gamma v^2}.
\label{h_qandq*}
\end{equation}
From (\ref{equivalence}), (\ref{h_qandq*}), and the third result stated above, we also obtain an expression of the joint conditional distribution function.
That is,
\begin{equation}
\frac{P(y_0,y,q,t|p_{\rm st}^{U,v^{\prime}\beta^{\prime}})}{P(q,t|p_{\rm st}^{U,v^{\prime}\beta^{\prime}})}
\sim p_{\rm can}^{U,\tilde v_{\rm i},\tilde \beta_{\rm i}}(y_0)p_{\rm can}^{U,\tilde v_{\rm f},\tilde \beta_{\rm f}}(y) 
+\frac{1}{1/t}o(1/t)
\label{condresult}
\end{equation}
for $y_0=Y_0L$, $y=YL$ $(-1<Y_0,Y<1)$ with
\begin{equation}
\tilde \beta_{\rm i}= \beta^{\prime} -\frac{q-\gamma v^2}{2T\gamma v^2},
\label{condbetaini}
\end{equation}
\begin{equation}
\tilde v_{\rm i}= \frac{v^{\prime}\beta^{\prime}+(q-\gamma v^2)/(2T\gamma v)}{\beta^{\prime}-(q-\gamma v^2)/(2T\gamma v^2)},
\end{equation}
\begin{equation}
\tilde \beta_{\rm f}= \beta  + \frac{q-\gamma v^2}{2T\gamma v^2},
\label{condbetafin}
\end{equation}
\begin{equation}
\tilde v_{\rm f}=v.
\end{equation}From (\ref{condresult}), (\ref{condbetaini}), and (\ref{condbetafin}), we find that
the particle
tends to climb up 
the potential and to reach the top of the potential
at time $t$ if $Q(t)/t$ is smaller than
$-\gamma v^2$
[or to climb down the potential from the top at time $0$
if $Q(t)/t$ is larger than $\gamma v^2(2\beta^{\prime}/\beta+1)$].
Here, we show that one can obtain the singularity
from these rare trajectories
by using an intuitive argument.

Now, let us imagine that we measure trajectories
and evaluate $G(0,h_q,t|p_{\rm st}^{U,v^{\prime},\beta^{\prime}})$
from the trajectories.
We set $t$ to be sufficiently large. Then,
from (\ref{scaledlegendre}), 
the trajectories required for the calculation
of $G(0,h_q,t|p_{\rm st}^{U,v^{\prime},\beta^{\prime}})$
must satisfy $Q(t)/t=q^*$, where $q^*$ was
given by (\ref{h_qandq*}).
Here, we consider the case that
$\beta_{\rm f}$
is negative. 
It indicates that the trajectories for the calculation
of $G(0,h_q,t|p_{\rm st}^{U,v^{\prime},\beta^{\prime}})$
also satisfy $y(t)=L$ because of the negative
inverse temperature.
Here, by using Jensen's inequality, we obtain
\begin{equation}
\begin{split}
&G(h_w,h_q,t|p_{\rm st}^{U,v^{\prime},\beta^{\prime}})\\
&\geq 
\frac{1}{t} \int _{0}^{t} dt
\left [h_q
\left \langle \dot Q(t) \right \rangle_{p_{\rm st}^{U,v^{\prime},\beta^{\prime}}} +h_w
\left \langle  \dot W(t) \right \rangle_{p_{\rm st}^{U,v^{\prime},\beta^{\prime}}}\right ]\\
&=- \frac{h_q}{t} 
\left \langle U(y(t))-U(y(0)) \right \rangle_{p_{\rm st}^{U,v^{\prime},\beta^{\prime}}}\\
&+\frac{h_w+h_q}{t} \int _{0}^{t} dt
\left \langle  \dot W(t) \right \rangle_{p_{\rm st}^{U,v^{\prime},\beta^{\prime}}},
\end{split}
\end{equation}
where we used (\ref{first_law_of_thermodynamics}) at the last line.
We evaluate the expectation value in the right-hand side by using the 
trajectories discussed above.
The first term is approximated as $-h_qU(L)/t$.
The second term can be omitted by assuming that 
the particle moves around the bottom
of the potential during most of the time, then suddenly climbs up the potential
just before the time $t$ and reaches the top of the potential at the time
$t$ [$y(t)=y$]. Thus, we have
\begin{equation}
G(h_w,h_q,t|p)\gtrsim -\frac{h_q}{t}U(L).
\end{equation}
This yields (\ref{Lfirst}).
We note that the case that $\beta_{\rm i}$ is negative
can also be discussed by following the same argument above.
The difference is that 
the particle
goes down to the bottom of the potential from the top
instead of climbing up.
From these arguments, we understand how hardly measurable trajectories
cause the singularity.

\section{Derivation}
\label{Sectionderivation}
Here, we derive the results.
In the first subsection, we analyze the
system with $L$ fixed.
Then, in the second subsection, we perform
a boundary layer analysis
by considering the limit $L\rightarrow \infty$.
Finally, in the third subsection, we derive the main results
of the paper.

\subsection{The method of the largest eigenvalue problem and
the Cole-Hopf transformation}
\label{SubsectionBiased}
We define an operator $\mathcal L_{h_w,h_q}^{(y)}$
by
\begin{equation}
\begin{split}
&\mathcal L^{(y)}_{h_w,h_q}\cdot \varphi \\
&= -\frac{\partial }{\partial y}\left [ \left (-\frac{1}{\gamma} \frac{\partial }{\partial y}U(y)+v \right ) \varphi \right ]
+h_wv\left ( \frac{\partial}{\partial y}U(y) \right ) \varphi \\
& +h_q\left [\frac{1}{\gamma} \left (\frac{\partial U(y)}{\partial y} \right )^2 
-\frac{T}{\gamma}\frac{\partial^2}{\partial y^2}U(y)
 \right ]\varphi +\frac{T}{\gamma}\frac{\partial ^2}{\partial y^2}\varphi 
 \\
& +\frac{T}{\gamma}\left (\frac{\partial U(y)}{\partial y} \right ) ^2(h_q)^2 \varphi  +\frac{2T}{\gamma}h_q
\frac{\partial }{\partial y} \left [ \frac{\partial U(y)}{\partial y}
\varphi \right ].
\end{split}
\label{l_forward}
\end{equation}
We denote the eigenfunctions of the operator
by $\psi_n$ $(n=0,1,2,...)$
and the corresponding eigenvalues by
$\mu_{n}$ $(n=0,1,2,...)$.
Here, the eigenvalues are labeled such that
${\rm Re}(\mu_n) \leq {\rm Re}(\mu_m)$ for $n>m$,
where Re$(a)$ is the real part of $a$.
We also consider the adjoint operator of $\mathcal L_{h_w,h_q}^{(y)}$, which is given by
\begin{equation}
\begin{split}
&\mathcal L^{(y)\dagger}_{h_w,h_q}\cdot \varphi \\
&= \left (-\frac{1}{\gamma} \frac{\partial }{\partial y}U(y)+v \right )
\frac{\partial }{\partial y} \varphi 
+h_wv\left ( \frac{\partial}{\partial y}U(y) \right ) \varphi \\
& +h_q\left [\frac{1}{\gamma} \left (\frac{\partial U(y)}{\partial y} \right )^2 
-\frac{T}{\gamma}\frac{\partial^2}{\partial y^2}U(y)
 \right ]\varphi +\frac{T}{\gamma}\frac{\partial ^2}{\partial y^2}\varphi 
 \\
& +\frac{T}{\gamma}\left (\frac{\partial U(y)}{\partial y} \right ) ^2(h_q)^2 \varphi  -\frac{2T}{\gamma}h_q \left ( \frac{\partial U(y)}{\partial y} \right )
\frac{\partial }{\partial y}  \varphi.
\end{split}
\label{Ldagger}
\end{equation}
We denote the eigenfunctions of $\mathcal L_{h_w,h_q}^{(y) \dagger}$ by
$\phi_n$ $(n=0,1,2,...)$ and the corresponding eigenvalues by
$\nu_{n}$ $(n=0,1,2,...)$. Generally, we may set
\begin{equation}
\nu_n=(\mu_{n})^{*}.
\end{equation}
The largest eigenvalues of $\mathcal L_{h_w,h_q}^{(y) \dagger}$
and $\mathcal L_{h_w,h_q}^{(y)}$ are real and do not
degenerate, which indicates
\begin{equation}
\nu_0=\mu_{0}.
\label{mu_nu}
\end{equation}
We also note that the eigenfunctions corresponding to the largest eigenvalue
are real. These results come from the Perron-Frobenius
theory. See the Appendix B of Ref. \cite{Nemoto-Sasa2}.
The orthonormal conditions for the eigenfunctions are
\begin{equation}
\int _{-L}^{L}dy (\phi_n(y))^*\psi_m(y)=\delta_{n,m}
\label{normalization_orthogonal}
\end{equation}
$(n,m=0,1,2,...)$, where $\delta_{n,m}$ is the Kronecker $\delta$.

Here, we define $q_{h_w,h_q}(y_0,y,t|p)$
by
\begin{equation}
q_{h_w,h_q}(y_0,y,t|p)=e^{tG(h_w,h_q,t|p)}P_{h_w,h_q}(y_0,y,t|p).
\label{definitionofq}
\end{equation}
As shown in Appendix \ref{Appendix_derivation1}, $\mathcal L_{h_w, h_q}^{(y)}$ turns out to be the
time evolution operator of $q_{h_w,h_q}(y_0,y,t|p)$. That is,
\begin{equation}
\frac{\partial }{\partial t}q_{h_w,h_q}(y_0,y,t|p)=
\mathcal L_{h_w, h_q}^{(y)}\cdot q_{h_w,h_q}(y_0,y,t|p).
\label{time_evolution_q}
\end{equation}
We expand $q_{h_w,h_q}(y_0,y,t|p)$ by the eigenfunctions $(\psi_n(y))_{n=0}^{\infty}$ and solve the time evolution equation (\ref{time_evolution_q})
with the initial condition $q_{h_w,h_q}(y_0,y,0|p)=p(y_0)\delta (y-y_0)$.
The result is
\begin{equation}
q_{h_w,h_q}(y_0,y,t|p)=p(y_0)\sum_{n=0}^{\infty} \left (\phi_{n}(y_0)\right )^*\psi_n(y) e^{\mu_n t}.
\label{expand_of_q_a}
\end{equation}
Then, we consider the large $t$ behavior of $q_{h_w,h_q}(y_0,y,t|p)$.
The $n=0$ term becomes dominant in the right-hand side of (\ref{expand_of_q_a}).  By combining the result with the definition (\ref{definitionofq}),
we obtain
\begin{equation}
\begin{split}
&P_{h_w,h_q}(y_0,y,t|p) \\
&=e^{\left ( \mu_0 -G(h_w,h_q,t|p)\right ) t}\\
&\times \left [ p(y_0) \phi_{0}(y_0)\psi_{0}(y) +
O\left  (e^{-(\mu_0-{\rm Re}(\mu_1))t }\right ) \right ].
\label{expand_of_q}
\end{split}
\end{equation}
Furthermore, by integrating (\ref{expand_of_q}) with respect to $y_0$ and $y$,
and taking the logarithm of it, we also obtain
\begin{equation}
\begin{split}
&G(h_w,h_q,t|p) \\
&= \mu_0 + \frac{1}{t}\log c_0\tilde c_0 + O\left  ( e^{-(\mu_0-{\rm Re}(\mu_1))t} \right ) ,
\label{qasymptotics}
\end{split}
\end{equation} 
where $c_0$ and $\tilde c_0$ are defined by
\begin{equation}
c_0=\int_{-L}^{L} dy \phi_{0}(y) p(y),
\label{c_n}
\end{equation}
\begin{equation}
\tilde c_{0}=\int_{-L}^{L} dy \psi_0(y).
\label{c_0tilde}
\end{equation}  
By comparing (\ref{asymptotic_expression_of_cumulant}) with (\ref{qasymptotics}), we arrive at
\begin{equation}
G_{\rm scaled}(h_w,h_q)=\mu_{0},
\label{G_mu}
\end{equation}
\begin{equation}
H_{\rm ex}(h_w,h_q|p)=\log c_0\tilde c_0.
\label{H_with_c_ctilde}
\end{equation}
Here, (\ref{G_mu}) is a well-known result \cite{Dembo,Touchette}.
There have been a lot of applications in which
(\ref{G_mu}) was used. See Ref. \cite{Lebowitz_Spohn}, for example.
The result (\ref{H_with_c_ctilde}) was used for the calculation
of an excess heat in Ref. \cite{Sagawa-Hayakawa}.
From (\ref{G_mu}) and (\ref{H_with_c_ctilde}),
we find that the expression (\ref{expand_of_q}) indicates that the 
biased joint distribution function becomes 
$p(y_0) \phi_{0}(y_0)\psi_0(y)e^{-H_{\rm ex}(h_w,h_q|p)}$
when $t$ is large.
Essentially the same result was discussed in Ref. \cite{Jack}.

Next, we use  the Cole-Hopf transformation in the 
largest eigenvalue problems $\mathcal L_{h_w,h_q}^{(y)}$
and $\mathcal L_{h_w,h_q}^{(y)\dagger}$,
and convert the largest eigenvalue problems to a non-linear
eigenvalue problem.
Here, we only see the results.
We define a non-linear operator $\mathcal M_{h,v}$ by
\begin{equation}
\begin{split}
\mathcal M_{h,v}\cdot \varphi
=& 2Thv \frac{\partial U}{\partial y} 
 +\frac{1}{2\gamma}\varphi^2\\
&+\left (-\frac{1}{\gamma} \frac{\partial U}{\partial y}+v \right )
\varphi+\frac{T}{\gamma}\frac{\partial }{\partial y}\varphi .
\label{M_H^UV}
\end{split}
\end{equation}
Then, we consider a non-linear eigenvalue problem
\begin{equation}
\mathcal M_{h,v}\cdot w_{h,v} = K_{h,v},
\label{nonlinear_eigenvalue}
\end{equation}
where the constant $K_{h,v}$ and the periodic function $w_{h,v}(y)$ are simultaneously determined from the boundary
condition $w_{h,v}(-L)=w_{h,v}(L)$ and
the normalization
condition
\begin{equation}
\int_{-L}^{L} dy w_{h,v}(y)=0.
\label{normalization}
\end{equation}
We introduce a potential function of
$w_{h,v}(y)$ by
\begin{equation}
V_{h,v}(y)=-\int_{0}^{y} dz w_{h,v}(z) + \rm const.
\label{V_vs_w}
\end{equation}
From these preparations, we can show following relations
\begin{equation}
\phi_{0}(y)=\frac{1}{C} \exp \left [h_qU(y)-\frac{V_{h_q+h_w,v}(y)}{2T}\right ],
\label{phi_0expression1}
\end{equation}
\begin{equation}
\psi_{0}(y)=\frac{1}{\tilde C} \exp \left [-\left (h_q+\beta \right )U(y)-\frac{V_{-\beta - h_q-h_w,-v}(y)}{2T}\right ],
\label{psi_0expression1}
\end{equation}
\begin{equation}
G_{\rm scaled}(h_w,h_q)=\frac{K_{h_q+h_w,v}}{2T},
\label{KandG}
\end{equation}
where the coefficients $(C)^*\tilde C$
are determined from the normalization condition (\ref{normalization_orthogonal}).
The derivation of these relations is shown in Appendix \ref{Appendix_Cole-Hopf}.
We note that the similar arguments were presented in Refs.  \cite{Nemoto-Sasa1, Nemoto-Sasa2}.

\subsection{Boundary layer analysis with large $L$ limit}
\label{Subsectionboundary_result}
Here, we evaluate the asymptotic behavior of $w_{h,v}(y)$ and $K_{h,v}$ 
when $L$ is large.
In this calculation, we use the condition (\ref{potential_condition}).

We use a boundary layer analysis \cite{boundarylayer} in the non-linear eigenvalue problem (\ref{nonlinear_eigenvalue}).
First, 
we define $\tilde w_{h,v}(Y)$ by
\begin{equation}
\tilde w_{h,v}(Y)\equiv  w_{h,v}(LY),
\end{equation}
where $-1\leq Y \leq 1$.  
By using $\tilde w_{h,v}(Y)$, we rewrite the left-hand side of (\ref{nonlinear_eigenvalue}) as
\begin{equation}
\begin{split}
& 2Thv\frac{\partial U(YL)}{\partial y}
+\frac{1}{2\gamma}(\tilde w_{h,v})^2\\
&+
\left (-\frac{1}{\gamma}
\frac{\partial U(YL)}{\partial y}
+v \right )
\tilde w_{h,v}+\frac{1}{L}\frac{T}{\gamma}\frac{\partial }{\partial Y}\tilde w_{h,v}.
\label{nonlinear_operator2}
\end{split}
\end{equation}
Now, we treat $L^{-1}$ as a perturbation parameter.
Since the coefficient in front of $\partial \tilde w_{h,v}/\partial Y$ is
$O(L^{-1})$,
we expect that there exists a certain region $I_{\rm b}$ in which
$\tilde w_{h,v}(Y)$ changes rapidly
to satisfy the periodic boundary condition and the normalization condition.
That is,  
\begin{equation}
\left | \frac{\partial \tilde w_{h,v}(Y)}{\partial Y} \right |  \gg \left | \tilde w_{h,v}(Y) \right |
\end{equation}
for $Y\in I_{\rm b}$. When the width of the region $I_{\rm b}$ becomes 0 in the limit
$L\rightarrow \infty$, the region is called the boundary layer.
We assume the existence of the boundary layer in this problem.
The basic strategy of the boundary layer analysis is the following:
(i) constructing the solutions inside the boundary layer (called inner solution) and
outside the boundary layer (called outer solution), and (ii) asymptotically
matching those solutions so that the continuity and the boundary conditions are
satisfied.
See Ref. \cite{boundarylayer} for more details.
In this paper, we consider only the outer solution and
obtain the leading order of $\tilde w_{h,v}(Y)$ by using some assumptions.
We state the result here. The derivation is shown in Appendix \ref{Subsection_boundarylayer_derivation}.

\begin{figure}[tbh]
\includegraphics[width=4cm]{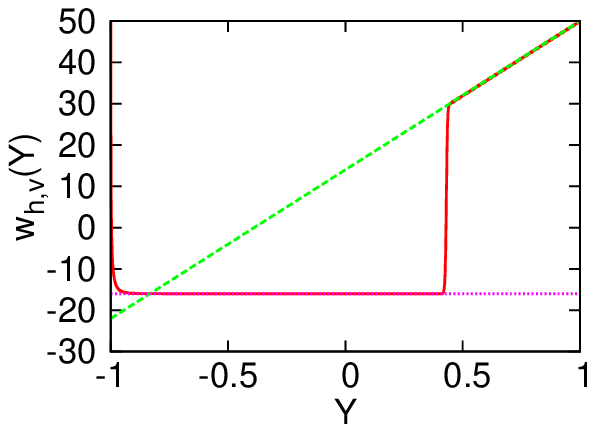}
\includegraphics[width=4cm]{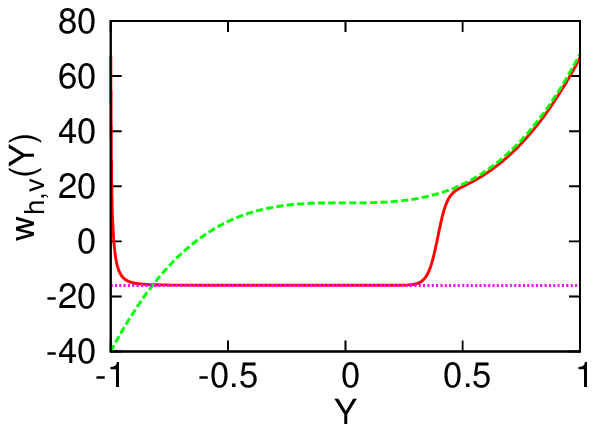}
\caption{(Color online) $\tilde w_{h,v}(Y)$ for potentials $U_{\rm harmo}(y)=(k/2)y^2$ (left) and $U_{\rm quart}(y)=(k_4/4)y^4$ (right).
Quantities are converted to dimensionless forms by setting 
$\gamma=T=v=1$. We fixed $k=1$, $k_4=1$ and $h=-8$.
We set $L=18$ for $U_{\rm harmo}(y)$ and $L=3$ for $U_{\rm quart}(y)$.
The green dashed lines and purple
dotted lines are obtained from (\ref{w_vh_posi}).
These lines correspond to $14+36Y$ and $-16$ for $U_{\rm harmo}(y)$,
and $14+54Y^3$ and $-16$ for $U_{\rm quart}(y)$.
The red solid lines are numerical results. 
It can be seen that (\ref{w_vh_posi}) agrees with the numerical results.
}
\label{fig1}
\end{figure}

When $L$ is large, by utilizing the condition (\ref{potential_condition}),
we obtain
\begin{equation}
\begin{split}
&\tilde w_{h,v}(Y) \\
&= \begin{cases}
2T\gamma hv   \quad  & -1\leq Y\leq a_{-}\\
 2\gamma \left [-v(1+Th)+(1/\gamma ) \partial U(YL) /\partial y
\right ] \quad & a_{-} \leq Y \leq 1
\end{cases}
\end{split}
\label{w_vh_posi}
\end{equation}
for $hv\leq 0$ and
\begin{equation}
\begin{split}
&\tilde w_{h,v}(Y) \\
&= \begin{cases}
 2\gamma \left [-v(1+Th)+(1/\gamma) \partial U(YL) /\partial y
\right ] \quad & -1 \leq Y \leq a_{+} \\
2T\gamma hv   \quad  & a_{+} \leq Y\leq  1
\end{cases}
\end{split}
\label{w_vh_minus}
\end{equation}
for $hv\geq 0$.
The coefficients $a_{+}$ and $a_{-}$ are determined
from the condition (\ref{normalization}). That is,
\begin{equation}
\gamma  v a_- \left (1+2Th \right ) -\gamma v
+\frac{U(L) -U(a_-L)}{L}=0
\label{a_-condition}
\end{equation}
and
\begin{equation}
\gamma  v a_+ \left (1+2Th \right ) +\gamma v
-\frac{U(La_+) -U(-L)}{L}=0.
\label{a_+condition}
\end{equation}
The examples of (\ref{w_vh_posi})
for the potentials $U_{\rm harmo}$ and
$U_{\rm quart}$ are displayed 
as green dashed lines and purple
dotted lines in Fig. \ref{fig1}.
We also evaluate $\tilde w_{h,v}(Y)$
by numerically solving (\ref{nonlinear_eigenvalue}).
The numerical method is the same as the one used in Ref. \cite{Nemoto-Sasa2}.
The obtained lines are displayed as red solid lines
in Fig. \ref{fig1}.
The figure shows that (\ref{w_vh_posi})
agrees with the numerical results.

\begin{figure}[tbh]
\includegraphics[width=4.0cm]{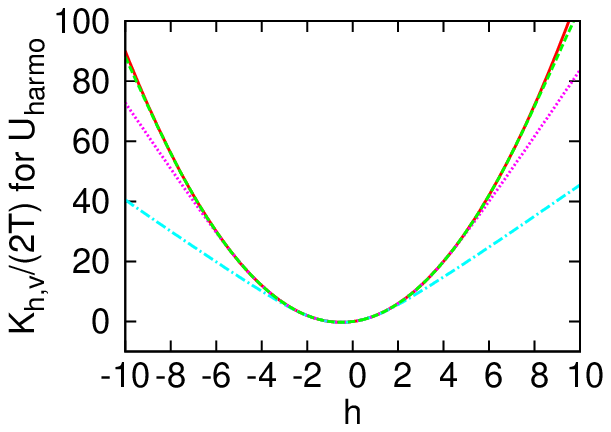}
\includegraphics[width=4.0cm]{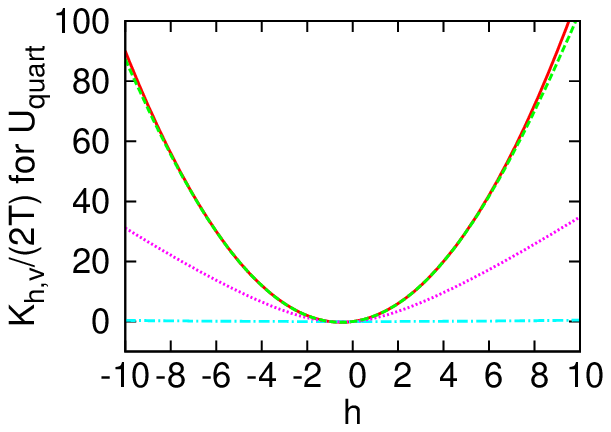}
\includegraphics[width=4.0cm]{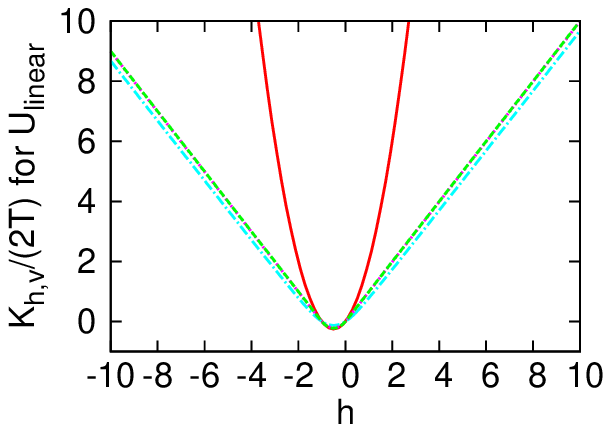}
\caption{(Color online)
$K_{h,v}/(2T)$ for potentials $U_{\rm harmo}(y)=(k/2)y^2$, $U_{\rm quart}(y)=(k_4/4)y^4$, and $U_{\rm linear}(y)=k_1|y|$.
Quantities are converted to dimensionless forms by setting 
$\gamma=T=v=1$. We fixed $k=1$, $k_4=1$ and $k_1=1$
and numerically evaluated $K_{h,v}/(2T)$.
In each figure, 
dashed dotted line (aqua blue), dotted line (purple), and dashed line
(green) correspond to
$L=6$, $L=12$ and $L=18$
for $U_{\rm harmo}$,
$L=1$, $L=2$ and $L=3$ 
for $U_{\rm quart}$, and
$L=5$, $L=25$ and $L=50$
for $U_{\rm linear}$.
The red solid lines in all figures denote $h+h^2$, which is predicted by (\ref{K_h__}).
As $L$ becomes large,
$K_{h,v}$ approaches
$h+h^2$
for $U_{\rm harmo}$ and $U_{\rm quart}$,
but does not for $U_{\rm linear}$.
Since $U_{\rm harmo}$ and $U_{\rm quart}$ satisfy
(\ref{potential_condition}) but
$U_{\rm linear}$ does not, these results are
consistent with our formulation.
}
\label{fig2}
\end{figure}

Next, since (\ref{nonlinear_eigenvalue}) is valid in the region that
$\tilde w_{h,v}(Y)=2T\gamma hv$, we obtain $K_{h,v}$ as
\begin{equation}
K_{h,v}=2T\left (\gamma v^2 h+ T\gamma  v^2 h^2\right ),
\label{K_h__}
\end{equation}
which leads to
\begin{equation}
G_{\rm scaled}(h_w,h_q)=\gamma v^2 (h_w+h_q)+ T\gamma  v^2 (h_w+h_q)^2.
\label{draggedcumulant}
\end{equation}
We also check (\ref{K_h__}) by comparing it with numerical results. See Fig \ref{fig2}.
It should be stressed that (\ref{draggedcumulant}) is always valid under
general periodic potentials,
as long as the condition (\ref{potential_condition}) is satisfied.


\subsection{Derivation of the main results of the paper}
From (\ref{a_-condition}) and (\ref{a_+condition}), we find
\begin{equation}
\lim_{L\rightarrow \infty}a_{\pm} =\mp 1,
\end{equation}
which indicates
\begin{equation}
\lim_{L\rightarrow \infty}\tilde w_{h,v}(Y)= 2T\gamma h v
\label{w_rightarrow}
\end{equation}
for fixed $Y$ ($-1<Y<1$).
From (\ref{w_rightarrow}), we calculate $V_{h_q+h_w,v}(y)$ and $V_{-\beta - h_q-h_w,-v}(y)$
as
\begin{equation}
\begin{split}
&\lim_{L\rightarrow \infty}V_{h_q+h_w,v}(YL)/L \\
&=-
\lim_{L\rightarrow \infty} \frac{1}{L}
\left [
\int _0^{YL}dzw_{h_q+h_w,v}(z) +{\rm const}.
\right ]\\
&= -2T\gamma v (h_q+h_w)Y +{\rm const}.
\label{dragged_V_opt}
\end{split}
\end{equation}
for fixed $Y$ $(-1<Y<1)$
and
\begin{equation}
\begin{split}
&\lim_{L\rightarrow \infty}V_{-\beta - h_q-h_w,-v}(y)/L \\
&=-\lim_{L\rightarrow \infty}\frac{1}{L}\left [ \int _0^{YL}dzw_{-\beta - h_q-h_w,-v}(z) +{\rm const}.\right ] \\
&= -2T\gamma v (\beta+h_q+h_w)Y+{\rm const}.
\end{split}
\label{dragged_Vtilde_opt}
\end{equation}
for fixed $Y$ $(-1<Y<1)$.
By substituting these expressions into (\ref{phi_0expression1})
and  (\ref{psi_0expression1}), we thus obtain $p(y_0)\phi_0(y_0)$
and $\psi_0(y)$ as
\begin{equation}
\begin{split}
&p(y_0)\phi_{0}(y_0) \\
&\sim  \frac{1}{C^{\prime}}\exp \left [h_qU(y_0) +\log p(y_0)+ \gamma v \left ( h_q+h_w \right )y_0\right ]
\label{phi_0expression2}
\end{split}
\end{equation}
for $y_0=Y_0L$ $(-1<Y_0<1)$
and
\begin{equation}
\psi_{0}(y)\sim  \frac{1}{\tilde C^{\prime}} \exp \left [-\left (\beta+h_q \right )U(y)
+\gamma v (\beta+h_q+h_w)y\right ]
\label{psi_0expression2}
\end{equation}
for $y=YL$ $(-1<Y<1)$,
where $C^{\prime}$ and $\tilde C^{\prime}$ are constants determined
by the normalization condition (\ref{normalization_orthogonal}).

Now, we derive the results in Section \ref{Sectionresult}.
We recall that the biased distribution function becomes
the original distribution function when we set $h_w=h_q=0$.
By substituting (\ref{psi_0expression2}) into (\ref{expand_of_q}),
we find that the stationary distribution
function of this system 
$p_{\rm st}^{U,v,\beta}(y)$ satisfies (\ref{result1}).

By recalling (\ref{draggedcumulant}), we obtain
(\ref{scaledlimit}).
Furthermore, because of (\ref{phi_0expression1})
and (\ref{psi_0expression1}), we have
\begin{equation}
\phi_0(y)=\psi_0(y)|_{(h_w,h_q,v)\rightarrow (-h_w, -\beta-h_q,  -v)}.
\end{equation}
Then, by utilizing (\ref{mu_nu}) and (\ref{G_mu}),
we obtain the fluctuation theorem for the scaled cumulant generating
function as
\begin{equation}
G_{\rm scaled}(h_w,h_q)=G_{\rm scaled}(-h_w,-\beta-h_q)|_{v\rightarrow -v}.
\end{equation}
This is (\ref{fluctuationtheorem}).

Here, we consider the case in which the initial distribution function
is equal to the stationary distribution function
$p_{\rm st}^{U,v^{\prime},\beta^{\prime}}(y)$. 
By substituting (\ref{phi_0expression2}) and (\ref{psi_0expression2})
into (\ref{expand_of_q}),
we obtain (\ref{biased_result_limit0}).

Finally, we derive (\ref{asymptoticH1}) and (\ref{asymptoticH2}).
By substituting (\ref{phi_0expression1}) and (\ref{psi_0expression1})
into (\ref{H_with_c_ctilde}), and 
noticing the normalization condition (\ref{normalization_orthogonal}),
we obtain
\begin{equation}
\begin{split}
&H_{\rm ex}(h_w,h_q|p)\\
&=\log  \int_{-L}^{L}dy \exp \left [h_qU(y)+\log p(y)-\frac{V_{h_q+h_w,v}(y)}{2T}\right ] \\
&+\log \int_{-L}^{L}dy\exp \left [-\left (h_q+\beta \right )U(y)-\frac{V_{-\beta - h_q-h_w,-v}(y)}{2T}\right ] \\
&-\log \int _{-L}^{L}dy\\
&\times \exp\left [-\beta U(y) -\frac{1}{2T}\left (V_{h_q+h_w,v}(y) + V_{-\beta - h_q-h_w,-v}(y)\right ) \right ].
\label{H_vs_potentials}
\end{split}
\end{equation}
By using (\ref{dragged_V_opt}) and (\ref{dragged_Vtilde_opt})
in this expression, we obtain
\begin{equation}
\begin{split}
H_{\rm ex}(h_w,h_q|p) \simeq  H + \tilde H + Z,
\end{split}
\end{equation}
where
\begin{equation}
\begin{split}
&H \\
&\equiv \log  \int_{-L}^{L}dy \exp \left [h_qU(y)+\log p(y) +  \gamma v \left ( h_q+h_w \right )y\right ],
\end{split}
\end{equation}
\begin{equation}
\begin{split}
&\tilde H  \\
&\equiv \log \int_{-L}^{L}dy\exp \left [-\left (h_q+\beta \right )U(y) + \gamma v(\beta + h_q +h_w)y\right ] ,
\end{split}
\end{equation}
and
\begin{equation}
Z\equiv -\log \int _{-L}^{L}dy\exp\left [-\beta U(y) +
\gamma v (\beta + 2h_q +2h_w) y \right ].
\end{equation}
$Z$ is always $O(1)$:
\begin{equation}
Z=O(1).
\label{Z_order}
\end{equation}
$\tilde H$ depends on the sign of $h_q+\beta$.
It becomes $O(1)$ for $h_q+\beta>0$, but $O(U(L))$
for $h_q+\beta<0$.
That is,
\begin{equation}
\tilde H = \begin{cases}
\mathcal  O(1) \quad &  h_q > - \beta   \\
O(U(L))  \quad & h_q <- \beta .
\end{cases}
\label{Htilde_order}
\end{equation}
$H$ depends on the initial distribution function $p(y)$.
In order to see it, we introduce a parameter $\beta_p$
\begin{equation}
\beta_p\equiv \lim_{L\rightarrow \infty}\frac{-\log p(L)}{U(L)},
\end{equation}
which can take a value from $0$ to $\infty$. It represents an effective 
temperature
of the initial distribution function.
By using $\beta_p$, we have
\begin{equation}
H = \begin{cases}
  O(U(L))  \quad & h_q > \beta_p \\
\mathcal  O(1) \quad &  h_q < \beta_p.
\end{cases}
\label{H_order}
\end{equation}
From (\ref{Z_order}), (\ref{Htilde_order}) and (\ref{H_order}), we arrive
at the conclusion
\begin{equation}
H_{\rm ex}(h_w,h_q|p)= \begin{cases}
O(U(L)) \quad &  h_q <- \beta   \\
O(1)  \quad &   -\beta <h_q < \beta_p   \\
O(U(L)) \quad &  h_q > \beta_p  .
\end{cases}
\end{equation}
By setting $p=p_{\rm st}^{U,v^{\prime},\beta^{\prime}}(y)$
in it,
we obtain (\ref{asymptoticH1}) and (\ref{asymptoticH2}).

\section{Concluding remarks}
\label{concluding_remarks}
In this paper, we studied the fluctuation of the work and the heat for a Brownian particle on
a moving periodic potential.
We considered a trapped particle limit
and discussed 
the Zon-Cohen singularity for an extended fluctuation theorem.
As the result,
we found that a conditional distribution function
given $Q(t)/t=q$, where $Q(t)/t$ was the rate of the heat dissipation,
became a canonical distribution function.
When $q$ was larger (or smaller) than a value,
the inverse temperature of the canonical distribution
function became negative.
This indicated that the particle climbed up (climbed down) the potential.
It turned out that this behavior of the particle
caused the Zon-Cohen singularity.

Before ending the paper, we touch on a 
possibility that the singularity takes place not only for the heat but also 
for the other quantities.
We showed that the singularity
occurred because of the non-interchangeability
of two types of limits.
The first one was the limit of large
observation time and 
the second one was the trapped particle limit.
These limits become non-interchangeable since
there exist rare trajectories in which
the particle reaches the top of the potential (or climbs down the potential
from the top).
Aside from the heat of trapped particle systems, 
there might exist a system in which the same kind
of singularity appears, if the system is defined by
a limit
and the limit is non-interchangeable with
the limit of large
observation time.
We would like to explore such a system for a deeper understanding
of fluctuation in nonequilibrium physics.

\section{Acknowledgement}
The author thanks S. Sasa for carefully reading this paper
and providing useful comments.
He also thanks S. Ito, K. Kawaguchi, M. Miyama, T. Sagawa and
H. Tasaki for related discussions.
This study was supported by a Grant-in-Aid for JSPS Fellows, No. 247538.

\appendix

\section{Derivation of (\ref{time_evolution_q})}
\label{Appendix_derivation1}
Here, we derive (\ref{time_evolution_q}).
We consider a joint distribution function of $y(0)$, $y(t)$, $W(t)$ and $Q(t)$,
which is defined by
\begin{equation}
\begin{split}
&P(y_0,y,W,Q,t|p)\\
&=p(y_0) \left \langle \delta (y(t) - y) \delta (W(t) -W) \delta(Q(t)-Q) \right \rangle_{y_0}.
\end{split}
\end{equation}
From the Langevin equations (\ref{y}), (\ref{W_y}), and (\ref{Q_y}), we have
the Fokker-Planck equation for $P(y_0,y,W,Q,t|p)$ as
\begin{equation}
\frac{\partial P}{\partial t} = \mathcal L_{\rm FP}^{(y,W,Q)}\cdot P,
\label{FPequation}
\end{equation}
where the Fokker-Planck operator $\mathcal L_{\rm FP}^{(y,W,Q)}$ is defined by
\begin{equation}
\begin{split}
&\mathcal L^{(y,W,Q)}_{\rm FP}\cdot \varphi \\
&= -\frac{\partial }{\partial y}\left [ \left (-\frac{1}{\gamma} \frac{\partial }{\partial y}U(y)+v \right ) \varphi \right ]
-v\left ( \frac{\partial}{\partial y}U(y) \right ) \frac{\partial}{\partial W}\varphi \\
& -\left [\frac{1}{\gamma} \left (\frac{\partial U(y)}{\partial y} \right )^2 
-\frac{T}{\gamma}\frac{\partial^2}{\partial y^2}U(y)
 \right ]\frac{\partial}{\partial Q}\varphi +\frac{T}{\gamma}\frac{\partial ^2}{\partial y^2}\varphi 
 \\
& +\frac{T}{\gamma}\left (\frac{\partial U(y)}{\partial y} \right ) ^2\frac{\partial ^2}{\partial Q^2} \varphi  -\frac{2T}{\gamma}
\frac{\partial ^2}{\partial Q\partial y} \left [ \frac{\partial U(y)}{\partial y}
\varphi \right ].
\end{split}
\label{FPoperator}
\end{equation}
By multiplying (\ref{FPoperator}) by $e^{Wh_w+Qh_q}$,
integrating it with respect
$W$ and $Q$, and noticing the definitions 
(\ref{biased_distribution}) and (\ref{definitionofq}), we obtain (\ref{time_evolution_q}).

\section{Derivation of (\ref{phi_0expression1}),
(\ref{psi_0expression1}), and (\ref{KandG})
with the Cole-Hopf transformation}
\label{Appendix_Cole-Hopf}
Here, we derive (\ref{phi_0expression1}),
(\ref{psi_0expression1}), and (\ref{KandG})
from the 
largest eigenvalue problems of $\mathcal L_{h_w,h_q}^{(y)}$
and $\mathcal L_{h_w,h_q}^{(y)\dagger}$.
Similar calculations were done in Refs. \cite{Nemoto-Sasa1, Nemoto-Sasa2}.

First, we consider the largest eigenvalue problem of $\mathcal L_{h_w,h_q}^{(y)\dagger}$,
\begin{equation}
\mathcal L_{h_w,h_q}^{(y)\dagger}\cdot\phi_0 = \nu_0 \phi_0.
\label{derivationLdagger}
\end{equation}
By dividing this by $\phi_0$ and performing some calculations, we obtain
\begin{equation}
\begin{split}
&\nu_0= (h_q+h_w)v \frac{\partial U}{\partial y} 
+\frac{T}{\gamma}\left [ \frac{\partial}{\partial y}\left (  \log \phi_0 -h_q U \right ) \right ]^2 \\
&+\left (-\frac{1}{\gamma} \frac{\partial }{\partial y}U+v \right )
\frac{\partial}{\partial y}\left (  \log \phi_0 -h_q U \right ) \\
&+\frac{T}{\gamma}\frac{\partial}{\partial y}\left [\frac{\partial}{\partial y}\left (  \log \phi_0 -h_q U \right )\right ] .
\label{variational_derivation1}
\end{split}
\end{equation}
Then, we introduce a potential function $V_{0}(y)$ by
\begin{equation}
V_{0}(y)=-2T\left ( \log \phi_0(y) -h_q U(y) \right ).
\label{dagger_phi_w_relation}
\end{equation}
This transformation is called the Cole-Hopf transformation.
By substituting (\ref{dagger_phi_w_relation}) into (\ref{variational_derivation1}) and
combining it with (\ref{mu_nu}) and (\ref{G_mu}), we obtain
\begin{equation}
\mathcal M_{h_w+h_q,v}\cdot 
\left (-\frac{\partial V_0}{\partial y} \right )  =2TG_{\rm scaled}(h_w,h_q),
\label{variational_derivation2}
\end{equation}

Next, we consider the largest eigenvalue problem of $\mathcal L_{h_w,h_q}^{(y)}$,
\begin{equation}
\mathcal L_{h_w,h_q}^{(y)}\cdot\psi_0 = \mu_0 \psi_0.
\end{equation}
We divide this by $\psi_0$. Then, after some calculations, we obtain
\begin{equation}
\begin{split}
& \mu_0 = -\left  ( \frac{1}{T}+ h_q+h_w \right )(-v)\frac{\partial U}{\partial y} \\
&+\frac{T}{\gamma}\left [ \frac{\partial}{\partial y}\left (\log \psi_0 + \left (h_q+\frac{1}{T} \right )U \right ) \right ]^2 \\
&+\left (-\frac{1}{\gamma} \frac{\partial U}{\partial y}-v \right )
\frac{\partial}{\partial y}\left [\log \psi_0 + \left (h_q+\frac{1}{T} \right )U \right ]\\
&+\frac{T}{\gamma}\frac{\partial}{\partial y}\left [\frac{\partial}{\partial y}\left (\log \psi_0+ \left (h_q+\frac{1}{T} \right )U \right ) \right ].
\label{variational_4}
\end{split}
\end{equation}
Thus, by defining
\begin{equation}
\tilde V_0(y)=-2T \left [ \log \psi_0(y) + \left (h_q+\frac{1}{T} \right )U(y)\right ],
\label{phi_w_relation}
\end{equation}
substituting it into (\ref{variational_4}), and
combining it with (\ref{G_mu}), we obtain
\begin{equation}
\mathcal M_{-\beta - h_w - h_q,-v}\cdot 
\left (-\frac{\partial \tilde V_0}{\partial y} \right )  =2TG_{\rm scaled}(h_w,h_q),
\label{variational_derivation3}
\end{equation}
Note that the sign of the velocity in
the left-hand side of (\ref{variational_derivation3}) is
reversed. 
This reflects a reversed protocol of
moving the periodic potential.

From (\ref{dagger_phi_w_relation}), (\ref{variational_derivation2}), (\ref{phi_w_relation}), and (\ref{variational_derivation3}),
we obtain
(\ref{phi_0expression1}),
(\ref{psi_0expression1}), and (\ref{KandG}).
We mention that the uniqueness of the solution of the
non-linear eigenvalue problem (\ref{nonlinear_eigenvalue}) is guaranteed by
the Perron-Frobenius theory, because (\ref{nonlinear_eigenvalue})
can be rewritten as the same form as (\ref{derivationLdagger}) by
following the same calculation from
(\ref{variational_derivation2}) to
(\ref{derivationLdagger}).

\section{Derivation of (\ref{w_vh_posi})
and (\ref{w_vh_minus}) by using
Boundary layer analysis}
\label{Subsection_boundarylayer_derivation}
Here, we derive (\ref{w_vh_posi})
and (\ref{w_vh_minus}).
We consider the outer solution $\tilde w_{h,v}^{{\rm o}}(Y)$ of (\ref{nonlinear_eigenvalue}), which
satisfies
\begin{equation}
\left | \frac{\partial \tilde w_{h,v}^{{\rm o}}(Y)}{\partial Y} \right |  \approx \left |
 \tilde w_{h,v}^{{\rm o}}(Y) \right |,
\label{outer_condition}
\end{equation}
where $\approx$ means that the left-hand side and the right-hand side
are of the same order of magnitude.
We recall that (\ref{nonlinear_eigenvalue}) 
is a quadratic equation for $\tilde w_{h,v}(Y)$.
Then, by solving this,
we obtain the
following identity:
\begin{equation}
\begin{split}
\tilde w_{h,v}(Y) &= \gamma \Bigg [-v+\frac{1}{\gamma }\frac{\partial U(YL)}{\partial y} \\
&\qquad  \pm \left |v(1+2Th)-\frac{1}{\gamma }\frac{\partial U(YL)}{\partial y}
 \right |
\sqrt{1+R(Y)}\Bigg ],
\end{split}
\label{w_koutousiki}
\end{equation}
where $R(Y)$ is defined by
\begin{equation}
\begin{split}
R(Y) \equiv &\frac{-v^24Th(1+Th)+2K_{h,v}/\gamma }{\left [v(1+2Th)-(1/\gamma) \partial U(YL)/\partial y \right ]^2}\\
&+\frac{-2T/(\gamma^2 L)\partial \tilde w_{h,v}/\partial Y}{\left [v(1+2Th)-(1/\gamma )\partial U(YL)/\partial y \right ]^2}.
\label{expression_R(Y)}
\end{split}
\end{equation}
Here, we set $\tilde w_{h,v}(Y)=\tilde w_{h,v}^{{\rm o}}(Y)$ in
the right-hand side of (\ref{expression_R(Y)}).
From (\ref{outer_condition}) and (\ref{potential_condition}),
we find that the second term of it is negligible. 
Then, we assume that $K_{h,v}$, which is equal to $2TG_{\rm scaled}(h_w,h_q)|_{h_w+h_q=h}$, is $O(1)$.
By combining this assumption with (\ref{potential_condition}), we also find that the first term of (\ref{expression_R(Y)}) is negligible.
Therefore, we omit the term $R(Y)$ in (\ref{w_koutousiki}).
The result leads to 
an expression of the outer solution $\tilde w_{h,v}^{{\rm o}}(Y)$,
\begin{equation}
\begin{split}
&\tilde w_{h,v}^{{\rm o}}(Y) = \begin{cases}
2T\gamma hv \\
 2\gamma \left [-v(1+Th)+(1/\gamma ) \partial U(YL) /\partial y
\right ].
\end{cases}
\end{split}
\end{equation}
Now, in order to connect these two solutions, we use following assumptions.
First, we may treat the boundary layer of this problem as a connection between these two
outer solutions.
Second, the number of the connecting points
should be minimized. Third, the one of the connecting points must be
$Y=\pm1$ \cite{Commentappendix}.
By combining these assumptions with the normalization condition
(\ref{normalization}), we can uniquely determine the solution $\tilde w_{h,v}(Y)$ as (\ref{w_vh_posi})
and (\ref{w_vh_minus}).


\begin{thebibliography}{99}
\bibitem{Commentappendix}
The location of the point might depend on where the derivative of the
potential has discontinuity. In the cases of $U=U_{\rm harmo}$ and
$U=U_{\rm quart}$, the derivatives of $U$
have discontinuities at $Y= \pm1$ and
the one of the connecting point is located at $Y= \pm1$.
\end{thebibliography}
\end{document}